\documentclass[aps,prl,floatfix,epsfig,twocolumn,showpacs,preprintnumbers]{revtex4}
\usepackage{amssymb}
\usepackage{graphicx}
\usepackage{amsmath}

\input{tcilatex}
\begin{document}

\title{The electronic structures and magnetic properties of perovskite\\
ruthenates from constrained orbital hybridization calculations}
\author{Xiangang Wan$^{1}$, Jian Zhou$^{2}$ and Jinming Dong$^{1}$}
\affiliation{$^{1}$National Laboratory of Solid State Microstructures and Department of
Physics, Nanjing University, Nanjing 210093, China\\
$^{2}$Department of Materials Science and Engineering, Nanjing University,
Nanjing, 210093 China}
\date{\today}

\begin{abstract}
We introduce a method to analyze the effect of hybridization by shifting
corresponding atomic levels using external potentials. Based on this
approach, we study perovskite ruthenates,\ and unambiguously identify that
the covalency between the \textit{A}-site cation and O ion will modify the
Ru-O hybridization and change the density of state at Fermi level,
consequently affect the magnetic properties significantly. We also study the
effect of pressure and reveal that hydrostatic pressure has a small effect
on the Ru-O-Ru bond angle of SrRuO$_{3}$, while it will decrease the Ru-O
length and increase the band width significantly. Therefore, the magnetic
ordering temperature will decrease monotonically with pressure.
\end{abstract}

\date{\today}
\maketitle

Due to the interesting electrical, structural and magnetic properties,
perovskite ruthenates have attracted continual research attention\cite%
{CaRuO3 PM, Exp, str SrRuO3, FM SrRuO3, FM CaRuO3,Local moment
picture,film,film2,BaRuO3,Chen-Zhou,Non-Fermi,Band 2,band 4,band 5, b 6,b 7,
Ru,Orbital ordering,MAE}. SrRuO$_{3}$ crystallizes in an orthorhombic
structure with space group \textit{Pbnm}\cite{str SrRuO3}. Same as SrRuO$_{3}
$, CaRuO$_{3}$\ also has an GdFeO$_{3}$-type perovskite\ structure\cite{str
SrRuO3}, the only observable difference between their crystal structures is
the Ru-O-Ru bond angle. However, their magnetic properties are different
completely. SrRuO$_{3}$\ is a ferromagnetic (FM) metal with a Curie
temperature T$_{c}$=160 K\cite{FM SrRuO3}, while there is a debate about the
ground state of CaRuO$_{3}$\cite{CaRuO3 PM,FM CaRuO3,Chen-Zhou,Non-Fermi}.
It is believed that the leading factor to determine the magnetic properties
of these 4\textit{d} materials is the structural distortion and the
associated Ru-O-Ru bond angle\cite{Band structure}. Comparing with Sr$^{2+}$%
, Ca$^{2+}$ has smaller radii, therefore CaRuO$_{3}$ has bigger structural
distortion and smaller Ru-O-Ru bond angle, which lifts the band degeneracy
and reduces the density of states at the Fermi energy N(E$_{f}$)\cite{Band
structure}. Consequently the Stoner criterion for magnetism may not be
satisfied for CaRuO$_{3}$. On the other hand, Ba$^{2+}$ has bigger radii, so
perovskite BaRuO$_{3}$ will have larger Ru-O-Ru bond angle, comparing with
that of SrRuO$_{3}$. Thus it is very natural to expect that BaRuO$_{3}$
should have higher T$_{c}$ than SrRuO$_{3}$.

Very recently, using high pressure and high temperature techniques, Jin 
\textit{et al.}\cite{Pressure BaRuO3}, synthesize the cubic perovskite BaRuO$%
_{3}$ successfully. The Ru-O-Ru bond angle of cubic BaRuO$_{3}$ is 180$^{%
\mathrm{o}}$, which is indeed larger than that of SrRuO$_{3}$. However, its T%
$_{c}$ is only 80 K, which clearly indicates that in addition to the bond
angle,\ there is other factors which also play important roles in the
magnetic properties. Jin \textit{et al.}\cite{Pressure BaRuO3} propose that
the weaker Ba-O hybridization will enhance the strength of Ru-O bond,
consequently reduce the T$_{c}$. The importance of \textit{A}-site cation
has also been emphasized by the density function calculations \cite{A-O
3d,A-O-2,A-O}. However contrast to Jin\cite{Pressure BaRuO3}, Zayak \textit{%
et al.}\cite{A-O} suggest that the magnetization in these compounds is
anticorrelated with the degree of covalency between \textit{A}-site ion and
O ion.

In addition to Ru-O-Ru bond angle and \textit{A}-site hybridization, Ru-O
bond length is another factor which also affect the band structure and the
N(E$_{f}$), consequently change the magnetic properties. Hydrostatic
pressure may vary both the bond length and angle, and there are several
publications which study the effect of pressure\cite{Pressure SrRuO3 2,
Pressure SrRuO3 Saturature, Pressure 2007, Pressure SrRuO3}. All experiments
indicate that the T$_{c}$ of SrRuO$_{3}$ will decrease with pressure\cite%
{Pressure SrRuO3 2, Pressure SrRuO3 Saturature, Pressure 2007, Pressure
SrRuO3}. However, the reason why T$_{c}$ decreases with pressure is still
not clear. X-ray diffraction\cite{Pressure 2007} suggests the Ru-O-Ru bond
angle will decrease with pressure, which results in the decrease of T$_{c}$.
On the other hand, neutron powder diffraction\cite{Pressure SrRuO3} claims
that the Ru-O bond length is sensitive to pressure while the bond angle
remains almost constant. Therefore, to clarify the controversial issue about
the influence of \textit{A}-O covalence and pressure on the magnetic
properties of perovskite ruthenates is an interesting problem which we
address in the present work.

Local spin density approximation (LSDA) is well used for perovskite
ruthenates\cite{Exp, Band 2, A-O-2, A-O, Band structure, LSDA is fine}. The
main aim of our study is to illustrate the effect A-O hybridization, we thus
neglect the effect of Coulomb correlation, although there is a debate about
whether the electronic correlation is important in perovskite ruthenates\cite%
{LSDA is fine,FM SrRuO3,band 3,Non-Fermi}. We perform the electronic
band-structure calculations\ within LSDA using the full potential
linearized-muffin-tin-orbital LMTO method\cite{LMTO}. A mesh of 256 k-points
in the irreducible Brillouin Zone is used. The self-consistent calculations
are considered to be converged when the difference in the total energy of
the crystal do not exceed 0.25 meV per formula unit (f.u.) and the
difference in the total electronic charge do not exceed 10$^{-3}$ electronic
charge at consecutive steps.

Based on the experimental structure\cite{Pressure BaRuO3}, we perform
calculation for the recently synthesized cubic BaRuO$_{3}$. We find that the
FM metallic solution is the ground state, which agrees with the experiment%
\cite{Pressure BaRuO3}. The energy difference (E$_{PM-FM}$) between total
energy of its PM and FM solution is 25.6 meV/f.u. The magnetic moment at Ru
site ($M_{Ru}$) of BaRuO$_{3}$ is about 0.76 $\mu _{B}$, which is also close
to the experimental value (about 0.80 $\mu _{B}$)\cite{Pressure BaRuO3}. Due
to the strongly hybridization with the Ru 4\textit{d} state, there is
considerable moment locating at O site. We also perform calculation for SrRuO%
$_{3}$ and CaRuO$_{3}$ using their experimental structure\cite{str SrRuO3}.
Consistent with the experimental and previous theoretical work, our
calculation also predict the ground state of SrRuO$_{3}$ is FM metal. The
obtained E$_{PM-FM}$ is 38.9 meV/f.u., which is slightly larger than that of
BaRuO$_{3}$. The magnetic moment mainly locates at Ru site (about 0.93 $\mu
_{B}$), which is also in good agreement with the previous works\cite{Band
structure}. On the other hand, for CaRuO$_{3}$, the FM state is unstable in
energy, which is consistent with other theoretical work\cite{Band structure}.

After reproducing the experimental results, we now try to explore the
controversy about the effect of \textit{A}-O\ covalence. It is widely
accepted that the alkali elements at the \textit{A}-site of \textit{A}BO$_{3}
$ perovskite are highly ionic, \textit{A}-site only contribute to the
distortion of lattice structure and should not affect the Ru-O hybridization
considerably. However, recently the effect of hybridization between A-site
cation and O had been emphasized by several groups\cite{Pressure BaRuO3, A-O
3d, A-O-2, A-O}, although there is debate about its exact effect. Thus to
gain a conclusive comment about the effect of A-O hybridization is important.

It is well known that the strength of hybridization between two orbits is
strongly depend on their energy difference. We thus introduce a external
potential in the Kohn-Sham equation to shift the orbital level consequently
control the hybridization:%
\begin{equation*}
(H_{KS}+|ilm\sigma \rangle V_{ext}\langle ilm\sigma |)\psi =E\psi 
\end{equation*}

where, $H_{KS}$\ is the Kohn-Sham potential, the basis $|ilm\sigma \rangle $%
\ is the orbit which we try to shift ($i$\ denotes the site,\ $n$\ the main
quantum number, $l$\ the orbital quantum number, $m$\ the magnetic quantum
number and $\sigma $\ the spin index), and $V_{ext}$\ is the magnitude we
shift. By this constrained calculation, we can directly study how various
orbitals and hybridizations between them affect the physical properties of
solids.\ Although similar in spirit, this constrained orbital-hybridization
method is different with the constrained LSDA calculation where the external
potential is applied to a particular orbital to constrain the orbital
occupation.

For perovskite ruthenates,\ the bands near the Fermi energy (E$_{f}$), which
determine the magnetic properties, are mainly contributed by Ru 4\textit{d}
and O 2\textit{p}, on the other hand the \textit{d} band of alkali elements
are several eV above E$_{f}$.\ Thus when we upshift the empty band of 
\textit{A}-site atom by a external potential, the effective hybridization
between this band and the O 2\textit{p} band will decrease, therefore we can
study the effect of \textit{A}-O hybridization directly by our constrained
orbital-hybridization method.

Based on the experimental crystal structure\cite{str SrRuO3, Pressure BaRuO3}%
, we perform constrained orbital-hybridization calculation, the numerical
results reveal that the position of Sr 4\textit{d} band affects not only the
magnetic moment but also the magnetization energy E$_{PM-FM}$. Upshifting Sr
4\textit{d} band by 9 eV and 20 eV will reduce $M_{Ru}$ and E$_{PM-FM}$ to
1.10 $\mu _{B}$, 20.2 meV; and 0.4 $\mu _{B}$, 7.5 meV, respectively. These
results indicate the strength of Sr-O hybridization is correlated with the
magnetization. Our constrained orbital-hybridization calculations also show
that upshifting Ba 5\textit{d} band has a similar effect as in SrRuO$_{3}$.
The position of Ca 3\textit{d} will also affect the band structure of CaRuO$%
_{3}$, but shifting Ca 3\textit{d} band cannot make the FM state stable.

To further understand the effect of the \textit{A}-O hybridization\textbf{\ }%
on magnetic properties, we perform constrained orbital-hybridization
calculation for PM state and shows the density of states (DOS) of BaRuO$_{3}$
in Fig.1. It is found that the energy position of Ba 5\textit{d} band has an
important effect on its electronic band structure. Conventional LSDA
calculation predicts that the e$_{g}$ band of Ru mainly locates at -7.6 to
-5.0 eV\ and -0.2 to 5.1 eV, while t$_{2g}$ band has two sharp peaks
centered at -0.1 and -5.2 eV, respectively, as shown in Fig.1(a). Upshifting
Ba 5\textit{d} band depresses the hybridization between Ba and O,
consequently varies the band structure around E$_{f}$\ considerably. As
shown in Fig.1(b), decreasing the Ba-O hybridization will widen the e$_{g}$
band of Ru slightly, while change the t$_{2g}$ band significantly, make the t%
$_{2g}$ to distribute in the whole range from -7.0 to 0.9 eV, and decrease
N(E$_{f}$) considerably. It is well accepted that \textit{A}RuO$_{3}$ is an
itinerant magnetism, and an large N(E$_{f}$) in PM\ calculation is crucial
for the magnetization. Therefore, according to the Stoner theory, the 
\textit{A}-O hybridization can affect the magnetic properties considerably.
We also perform constrained orbital-hybridization calculation for SrRuO$_{3}$
and CaRuO$_{3}$. The similar sensitivity of electronic structure to energy
position of \textit{d} band of cation has also been found. So we can
conclude that the \textit{A}-O\ hybridization will change the Ru-O
hybridization, consequently change the magnetization of perovskite
ruthenates.

\begin{figure}[tbp]
\includegraphics[width=3.0in]{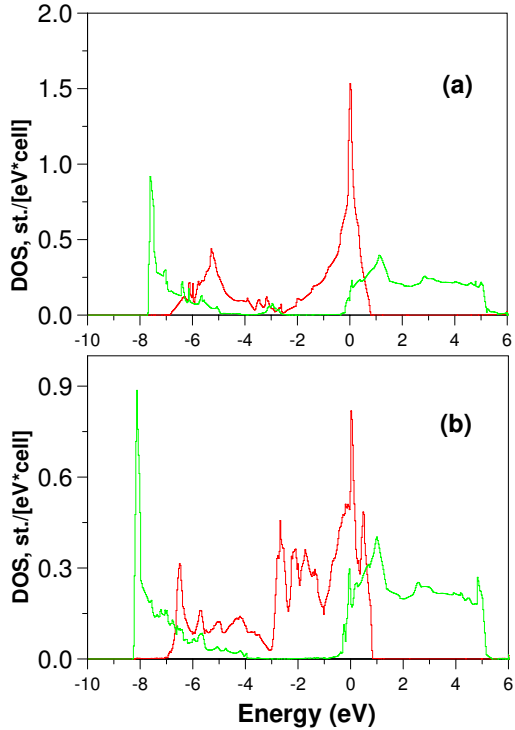}
\caption{(Color online). Partial density of states (DOS) of the cubic BaRuO$%
_{3}$. Red and green curve denote the t$_{2g}$ and e$_{g}$ band,
respectively. (a) is the conventional LSDA results; (b) is the constrained
orbital hybridization calculations results with the 5\textit{d} band of Ba
upshifted by 9 eV.}
\label{fig1}
\end{figure}

\begin{figure}[tbp]
\includegraphics[width=2.5in]{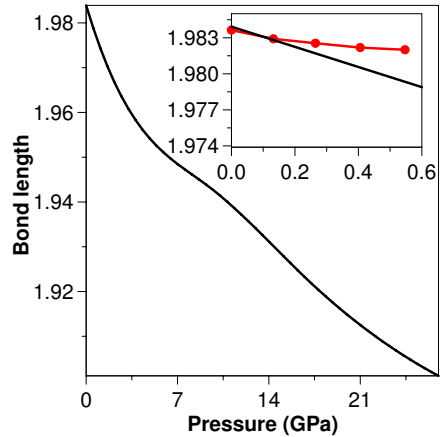}
\caption{(Color online). Numerical pressure dependence of the average Ru-O
bond length of SrRuO$_{3}$ (in a unit of $\mathring{A}$). The inset is a
comparison between the experimental\protect\cite{Pressure SrRuO3} (red) and
numerical (black) results at low pressures.}
\label{fig2}
\end{figure}

\begin{figure}[tbp]
\includegraphics[width=2.5in]{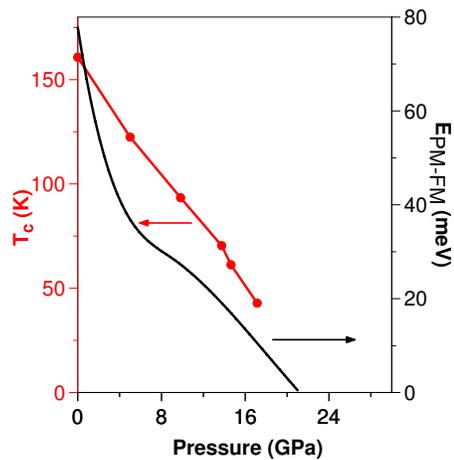}
\caption{(Color online). The numerical pressure dependence of the energy
difference between PM and FM solution of SrRuO$_{3}\ $(black), and the
experimental\protect\cite{Pressure 2007} pressure dependence of the T$_{c}\ $
(red).}
\label{fig3}
\end{figure}

To study the effect of pressure on the structure, we optimize the structure
and relax all independent internal atomic coordinates until the
corresponding forces are less than 1 mRy/a.u. Consistent with the recent
experiment\cite{Pressure SrRuO3}, we find that the Ru-O-Ru bond angle of
SrRuO$_{3}$ is not sensitive to the external pressure, while the Ru-O bond
length decreases monotonically with pressure, as shown in Fig.2. Our
numerical results agree with the experimental value at low pressures, as
shown in the inset of Fig.2, while the experimental bond length at high
pressure is not available. Based on the optimized structure, we perform the
band-structure calculation for a number of different volumes and fit the
curves E(V) of the calculated total energies vs volume to the
Birch-Murnaghan equation of state. The obtained equilibrium volume V$_{0}$,
bulk modulus at equilibrium B$_{0}$, its pressure derivative B$%
_{0}^{^{\prime }}$ for SrRuO$_{3}$, together with the available experimental
and theoretical values are summarized in Table I. Our theoretical
equilibrium volume is only about 1.7\% smaller than the experimental values.
Such a deviation exists normally in the LSDA calculations. Our full
potential calculations agree with the experiment very well, while the
pseudopotential calculation\cite{band 3} slightly overestimates the bulk
modulus.

\begin{table}[tbp]
\caption{The obtained equilibrium volume V$_{0}$, bulk modulus B$_{0}$ and
its pressure derivative B$_{0}^{^{\prime }}$ of SrRuO$_{3}$. Previous
experimental\protect\cite{Pressure 2007} and the theoretical results%
\protect\cite{band 3} from VASP and SIESTA package are also listed for
comparison.}%
\begin{tabular}{llll}
\hline
& V$_{0}$(a.u.$^{3}$) & B$_{0}$ (GPa) & B$_{0}^{\prime }$ \\ \hline
Present LMTO & 1605 & 190.5 & 4.6 \\ 
Experiment\cite{Pressure 2007} & 1633 & 192 & 5.3 \\ 
Other theory (VASP)\cite{band 3} &  & 200 & 4.6 \\ 
Other theory (SIESTA)\cite{band 3} &  & 219 & 4.4 \\ \hline
\end{tabular}%
\end{table}

Perovskite ruthenate belongs to itinerant magnet, therefore one cannot
estimate the interatomic exchange interaction and T$_{c}$ accurately as in
the local magnet\cite{J and Tc}. One important parameter relevant to the T$%
_{c}$ is E$_{PM-FM}$, which is plotted in Fig.3. For comparison, we also
plot the experimental curve of T$_{c}$ vs pressure in Fig.3. It is
interesting to notice that both numerical E$_{PM-FM}$ and experimental T$_{c}
$ decrease monotonically and rapidly. Theoretical work predicts that the
critical pressure (P$_{c}$) for FM to PM\ transition is about 21 GPa, after
this pressure E$_{PM-FM}$\ starts to equal or less than zero. The
experimental T$_{c}$ at the pressures above 18 GPa is not available, but an
extrapolation of the experimental data at low pressures yields a crude
approximation of 23 GPa, which is quite close to our numerical results.

In summary, we introduce a method to identify the effect of covalence for a
particular material by applying orbitally dependent external potential. This
approach allows us to control the hybridization between the orbits directly.
Our constrained orbital hybridization calculation unambiguously identify
that the covalency between the \textit{A}-site cation and O ion will modify
the Ru-O\ hybridization, consequently affect the magnetic properties
significantly. We also study the effect of pressure and find that
hydrostatic pressure has a small effect on the Ru-O-Ru bond angle of SrRuO$%
_{3}$, while it will decrease the Ru-O length, increase the band width and
result in the monotonic decrease of the magnetic ordering temperature.

We acknowledges support from National Key Project for Basic Researches of
China (No. 2006CB921802 and 2010CB923404), Natural Science Foundation of
China (No. 10774067 and 10974082) and Fok Ying Tung Education Foundation for
the financial support through Contract No. 114010. JZ was also supported by
the National Laboratory of Solid State Microstructures (2010YJ07).

\end{document}